\documentclass[twocolumn]{webofc}

\usepackage{color}
\usepackage[varg]{txfonts}   
\usepackage[mathscr,scaled=1.15]{urwchancal}
\DeclareFontFamily{OT1}{pzc}{}
\DeclareFontShape{OT1}{pzc}{m}{it}%
{<-> s * [1.15] pzcmi7t}{}
\DeclareMathAlphabet{\mathpzc}{OT1}{pzc}{m}{it}

\definecolor{purple}{rgb}{0.5,0,0.5}
\definecolor{blue}{rgb}{0.0,0,0.9}
\definecolor{prdblue}{rgb}{0.133,0.118,0.498}
\usepackage[colorlinks=true, pdfstartview=FitV, linkcolor=prdblue, citecolor= prdblue, urlcolor=prdblue]{hyperref}

\begin{document}
\title{$\,$\\[-6ex]\hspace*{\fill}{\normalsize{\sf\emph{Preprint no}. NJU-INP 004/19}}\\[1ex]
Resonance Electroproduction and the Origin of Mass}
%
%

\author{\firstname{Craig D.} \lastname{Roberts}\inst{1,2}\fnsep\thanks{\email{cdroberts@nju.edu.cn}}
}

\institute{School of Physics, Nanjing University, Nanjing, Jiangsu 210093, China
\and
           Institute for Nonperturbative Physics, Nanjing University, Nanjing, Jiangsu 210093, China
          }

\abstract{%
One of the greatest challenges within the Standard Model is to discover the source of visible mass.  Indeed, this is the focus of a ``Millennium Problem'', posed by the Clay Mathematics Institute.  The answer is hidden within quantum chromodynamics (QCD); and it is probable that revealing the origin of mass will also explain the nature of confinement.  In connection with these issues, this perspective describes insights that have recently been drawn using contemporary methods for solving the continuum bound-state problem in relativistic quantum field theory and how they have been informed and enabled by modern experiments on nucleon-resonance electroproduction.}
\maketitle
\section{Emergence of Mass}
\label{intro}
The natural energy-scale for strong interactions is characterised by the proton mass:
\begin{equation}
m_p \approx 1\,{\rm GeV} \approx 2000\,m_e\,,
\end{equation}
where $m_e$ is the mass of the electron.  In the Standard Model, $m_e$ is rightly attributed to the Higgs boson; but what is the source of the enormous enhancement to produce $m_p$?  This is the crux: the source of the vast majority of visible mass in the Universe is unknown.  Followed logically to its origin, this question leads to an appreciation that the existence of our Universe depends critically on, \emph{inter alia}, the following empirical facts:
(\emph{i}) the proton is massive, \emph{i.e}.\ the mass-scale for strong interactions is vastly different to that of electromagnetism;
(\emph{ii}) the proton is absolutely stable, despite being a composite object constituted from three valence-quarks;
and (\emph{iii}) the pion, responsible for long-range interactions between nucleons, is unnaturally light (not massless), possessing a lepton-like mass despite being a strongly interacting composite object built from a valence-quark and valence-antiquark.
These are basic emergent features of Nature; and QCD must somehow explain them and many other high-level phenomena with enormous apparent complexity.

The Lagrangian of chromodynamics is simple:
\begin{subequations}
\label{QCDdefine}
\begin{align}
{\mathpzc L}_{\rm QCD} & = \sum_{j=u,d,s,\ldots}
\bar{q}_j [i\gamma^\mu D_{\mu} - m_j] q_j - \tfrac{1}{4} G^a_{\mu\nu} G^{a\mu\nu},\\
\label{gluonSI}
D_{\mu} & = \partial_\mu + i g \tfrac{1}{2} \lambda^a A^a_\mu\,,\\
G^a_{\mu\nu}  & = \partial_\mu A^a_\nu + \partial_\nu A^a_\mu +
\underline{\textcolor[rgb]{0.00,0.0,00}{i g f^{abc}A^b_\mu A^c_\nu}}.
\label{gluonSIU}
\end{align}
\end{subequations}
Here: $\{q_j\}$ are the quark fields, with $j$ their flavor label and $m_j$ their Higgs-generated current-quark masses; and $\{ A_\mu^a, a=1,\ldots, 8 \}$ are the gluon fields, with $\{\tfrac{1}{2} \lambda^a\}$ the generators of the SU$(3)$ (color/chromo) gauge-group in the fundamental representation.  In comparison with quantum electrodynamics (QED), the \emph{solitary} difference is the term describing gluon self-interactions, marked as the underlined piece in Eq.\,\eqref{gluonSIU}.  Somehow, ${\mathpzc L}_{\rm QCD}$ -- this one line, along with two definitions -- is responsible for the origin, mass, and size of almost all visible matter in the Universe.  That being true, then QCD is quite possibly the most remarkable fundamental theory ever invented.

The only apparent mass-scales in Eq.\,\eqref{QCDdefine} are the current-quark masses, generated by the Higgs boson; but focusing on the valence quarks which define nucleons, \emph{i.e}.\ $u$, $d$, this scale is more-than 100-times smaller than $m_p$.  No amount of ``staring'' at ${\mathpzc L}_{\rm QCD}$ can reveal the source of that enormous amount of ``missing mass''; yet, it is there.  This contrasts starkly with QED wherein, \emph{e.g}.\ the scale in the spectrum of the hydrogen atom is set by $m_e$, a prominent feature of ${\mathpzc L}_{\rm QED}$ that is generated by the Higgs boson.

The emergence of mass is seen even more keenly by considering that even treated as a classical theory, chromodynamics is a non-Abelian local gauge field theory.  As with all such theories formulated in four spacetime dimensions, no energy-scale exists in the absence of Lagrangian masses for the matter fields.  (The absence of such masses defines the chiral limit.)  There is no dynamics in a scale-invariant theory, only kinematics: the theory looks the same at all length-scales; hence, there can be no clumps of anything.  Bound states are therefore impossible and, accordingly, our Universe cannot exist.

A \emph{spontaneous} breaking of symmetry, as realised via the Higgs mechanism, does not solve this problem: the masses of the neutron and proton, the kernels of all visible matter, are roughly 100-times larger than the Higgs-generated current-masses of the light $u$- and $d$-quarks, the main building blocks of protons and neutrons.

Consequently, the questions of how does a mass-scale appear and why does it have the value we observe are inseparable from the question of how did the Universe come into being.

Modern quantum field theories are built upon Poincar\'e invariance, so consider the energy-momentum tensor in classical chromodynamics, $T_{\mu\nu}$.  Conservation of energy and momentum is a consequence of spacetime translational invariance, elements in the set of Poincar\'e transformations; hence, $\partial_\mu T_{\mu\nu} = 0$.  Suppose now that one performs a global scale transformation on the coordinates and fields in classical chromodynamics.  In the absence of fermion masses, \emph{viz}.\ suppressing the Higgs mechanism, the classical action is invariant under the scale transformation and the dilation current is conserved:
\begin{equation}
{\cal D}_\mu = T_{\mu\nu} x_\nu \; | \;
\partial_\mu {\cal D}_\mu  = 0  = [\partial_\mu T_{\mu\nu} ] x_\nu + T_{\mu\nu} \delta_{\mu\nu}  = T_{\mu\mu}\,.
\label{SIcQCD}
\end{equation}
Evidently, the energy-momentum tensor must be traceless in a scale invariant theory.

Classical chromodynamics is not a meaningful framework for many reasons; amongst them the fact that strong interactions are empirically known to be characterised by a large mass-scale, $m_p \approx 1\,$GeV.  In quantising the theory, regularisation and renormalisation of ultraviolet divergences introduces a mass-scale.  This is ``dimensional transmutation'': all quantities, including the field operators themselves, become dependent on a mass-scale.  It entails violation of Eq.\,\eqref{SIcQCD}, \emph{i.e}.\ the appearance of the chiral-limit ``trace anomaly'':
\begin{equation}
\label{SIQCD}
T_{\mu\mu} = \beta(\alpha(\zeta))  \tfrac{1}{4} G^{a}_{\mu\nu}G^{a}_{\mu\nu} =: \Theta_0 \,,
\end{equation}
where $\beta(\alpha(\zeta))$ is QCD's $\beta$-function, $\alpha(\zeta)$ is the associated running-coupling, and $\zeta$ is the renormalisation scale.  Eq.\,\eqref{SIQCD} indicates that a mass-scale related to the resolving power of a given measurement is introduced via quantisation, \emph{viz}.\ the scale \emph{emerges} as an integral part of the theory's quantum definition.

There is another aspect of chromodynamics that should be highlighted; namely, the classical Lagrangian still defines a non-Abelian local gauge theory.  Accordingly, the concept of local gauge invariance persists; but without a mass-scale, there is no notion of confinement.  For example, three quarks can be prepared in a colour-singlet combination and colour rotations will keep the three-body system neutral; but the quarks involved need not have any proximity to one another.  Indeed, proximity is meaningless because all lengths are equivalent in a scale invariant theory.  Hence, the question of ``Whence a mass-scale?'' is equivalent to ``Whence a confinement scale?''.

Knowing a trace anomaly exists does not actually deliver a great deal: it only indicates that there is a mass-scale.  The crucial issue is whether one can compute and/or understand the magnitude of that scale.

One can certainly measure the size of the scale anomaly, for consider the expectation value of the energy-momentum tensor in the proton:
\begin{equation}
\label{EPTproton}
\langle p(P) | T_{\mu\nu} | p(P) \rangle = - P_\mu P_\nu\,,
\end{equation}
where the right-hand-side follows from the equations-of-motion for a one-particle proton state.  In the chiral limit,
\begin{align}
\label{anomalyproton}
\langle p(P) | T_{\mu\mu} | p(P) \rangle  = - P^2 & = m_p^2
= \langle p(P) |  \Theta_0 | p(P) \rangle\,;
\end{align}
namely, there is a clear sense in which it is possible to conclude that the entirety of the proton mass is produced by gluons.  The trace anomaly is measurably large; and that property must logically owe to gluon self-interactions, which are also responsible for asymptotic freedom.

This is a valid conclusion.  After all, what else could be responsible for a mass-scale in QCD?  QCD is all about gluon self-interactions; and it's gluon self-interactions that (potentially) enable one to rigorously (nonperturbatively) define the expectation value in Eq.\,\eqref{anomalyproton}.  On the other hand, it's only a sensible conclusion when the operator and the wave function are defined at a resolving-scale $\zeta \gg m_p$, \emph{viz}.\ when one employs a parton-model basis.

There is also another issue, which can be exposed by returning to Eq.\,\eqref{EPTproton} and replacing the proton by the pion:
\begin{equation}
\label{EPTpion}
\langle \pi(q) | T_{\mu\nu} | \pi(q) \rangle = - q_\mu q_\nu\,
\stackrel{\rm chiral\,limit}{\Rightarrow} \langle \pi(q) |  \Theta_0 | \pi (q) \rangle = 0
\end{equation}
because the pion is a massless Nambu-Goldstone (NG) mode.  Equation\,\eqref{EPTpion} could mean that the scale anomaly vanishes trivially in the pion state, \emph{viz}.\ that gluons and their self-interactions have no impact within a pion because each term in the practical computation of the operator expectation value vanishes when evaluated in the pion. However, that is a difficult way to achieve Eq.\,\eqref{EPTpion}.  It is easier to imagine that Eq.\,\eqref{EPTpion} owes to cancellations between different operator-component contributions.  Of course, such precise cancellation should not be an accident.  It could only arise naturally because of some symmetry and/or symmetry-breaking pattern.

Equations\,\eqref{anomalyproton} and \eqref{EPTpion} present a quandary, which highlights that no understanding of the source of the proton's mass can be complete unless it simultaneously explains the meaning of Eq.\,\eqref{EPTpion}.  Moreover, any discussion of confinement, fundamental to the proton's absolute stability, is impossible before this conundrum is resolved.  The explanation of these features of Nature must lie in the dynamics responsible for the emergence of $m_p$ as the natural mass-scale for nuclear physics; and one of the most important goals in modern science is to explain and elucidate the entire array of empirical consequences of this dynamics.

\section{Gluons are Massive}
\label{SecGluonMass}
Gluons must be the key; after all, their self interactions separate QCD from QED.  Gluons are supposed to be massless.  This is true in perturbation theory; but it is a feature that is not preserved nonperturbatively.  Beginning with a pioneering effort almost forty years ago \cite{Cornwall:1981zr}, continuum and lattice studies of QCD's gauge sector have been increasing in sophistication and reliability; and today it is known that the gluon propagator saturates at infrared momenta \cite{Boucaud:2011ug, Strauss:2012dg, Binosi:2014aea, Aguilar:2015bud, Binosi:2016nme, Gao:2017uox, Cyrol:2017ewj, Rodriguez-Quintero:2018wma, Binosi:2019ecz}:
\begin{equation}
\label{eqGluonMass}
\Delta(k^2\simeq 0) = 1/m_g^2.
\end{equation}
Thus, the long-range propagation characteristics of gluons are dramatically affected by their self-interactions.  Importantly, one may associate a renormalisation-group-invariant gluon mass-scale with this effect: $m_0 \approx 0.5\,$GeV\,$\approx m_p/2$, and summarise a large body of work by stating that gluons, although acting as massless degrees-of-freedom on the perturbative domain, actually possess a running mass, whose value at infrared momenta is characterised by $m_0$.

Asymptotic freedom ensures that QCD's ultraviolet behaviour is controllable; but the emergence of a gluon mass reveals a new frontier within the Standard Model because the existence of a running gluon mass, large at infrared momenta, has an impact on all analyses of the bound-state problem.  For instance, it could be a harbinger of gluon saturation \cite{Accardi:2012qut, NAP25171X}.
Furthermore, $m_0>0 $ entails that QCD dynamically generates its own infrared cutoff, so that gluons with wavelengths $\lambda \gtrsim \sigma :=1/m_0 \approx 0.5\,$fm decouple from the strong interaction, hinting at a dynamical realisation of confinement \cite{Brodsky:2015aia}.

\begin{figure}[t]
\centerline{\includegraphics[width=0.45\textwidth]{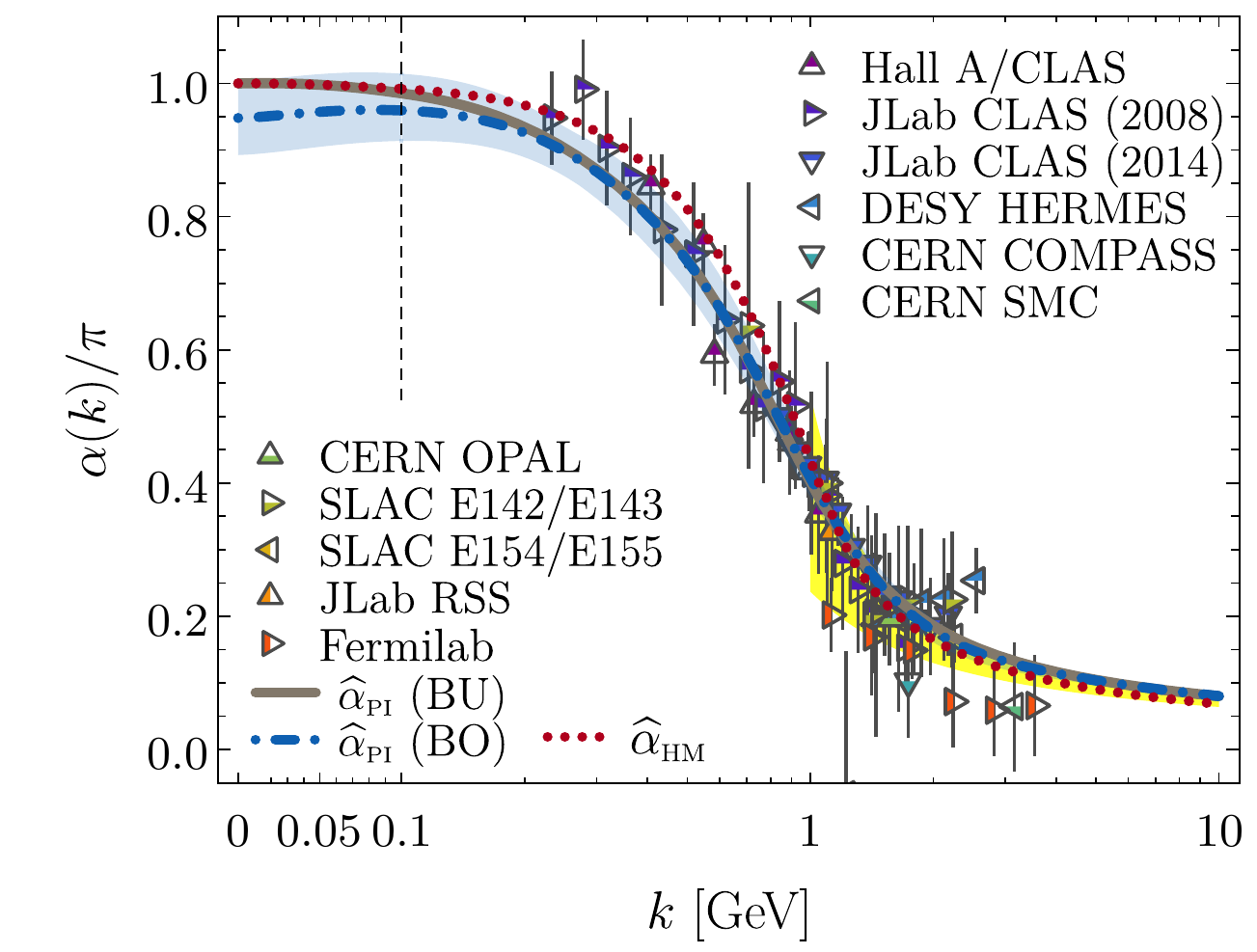}}
\caption{\label{FigwidehatalphaII}
Dot-dashed [blue] curve: process-in\-de\-pen\-dent running-coupling $\alpha_{\rm PI}(k^2)$ \cite{Binosi:2016nme}: the shaded [blue] band bracketing this curve combines a 95\% confidence-level window based on existing lQCD results for the gluon two-point function with an error of 10\% in the continuum analysis of relevant ghost-gluon dynamics.
Solid [black] curve, updated result \cite{Rodriguez-Quintero:2018wma}.
Also depicted, world's data on the process-dependent coupling $\alpha_{g_1}$, defined via the Bjorken sum rule, the sources of which are listed elsewhere \cite{Binosi:2016nme}.
The shaded [yellow] band on $k>1\,$GeV represents $\alpha_{g_1}$ obtained from the Bjorken sum by using QCD evolution to extrapolate high-$k^2$ data into the depicted region \cite{Deur:2008rf}; and, for additional context, the dashed [red] curve is the effective charge obtained in a light-front holographic model \cite{Deur:2016tte}.
The k-axis scale is linear to the left of the vertical dashed line and logarithmic otherwise.
}
\end{figure}

\section{Process-Independent Effective Charge}
There are many other consequences of the intricate nonperturbative nature of QCD's gauge-sector dynamics.  Important amongst them is the generation of a process-independent running coupling, $\alpha_{\rm PI}(k^2)$ -- see Refs.\,\cite{Binosi:2016nme, Rodriguez-Quintero:2018wma}.  Depicted as the solid [black] curve in Fig.\,\ref{FigwidehatalphaII}, this is a new type of effective charge, which is an analogue of the Gell-Mann--Low effective coupling in QED because it is completely determined by the gauge-boson propagator.  The result in Fig.\,\ref{FigwidehatalphaII} is a parameter-free prediction, capitalising on analyses of QCD's gauge sector undertaken using continuum methods and informed by numerical simulations of lattice-QCD (lQCD).

As a unique process-independent effective charge, $\alpha_{\rm PI}$ appears in every one of QCD's dynamical equations of motion, including the gap equation, setting the strength of all interactions.  $\alpha_{\rm PI}$ therefore plays a crucial role in understanding the dynamical origin of light-quark masses in the Standard Model even in the absence of a Higgs coupling, as described in the next section.

\begin{figure}[t]
\centerline{\includegraphics[clip,width=0.43\textwidth]{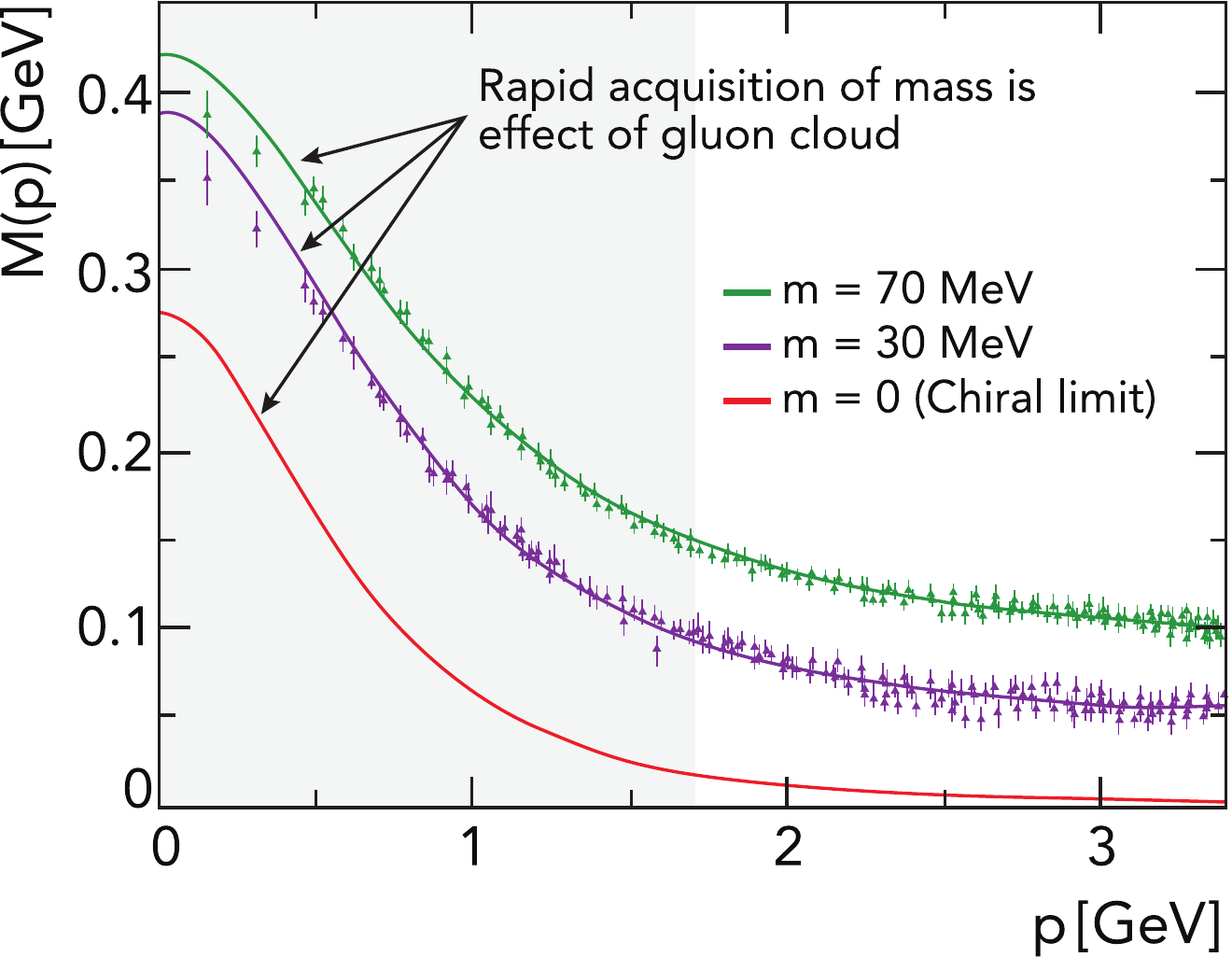}}

\caption{\label{gluoncloud}
Renormalisation-group-invariant dressed-quark mass function, $M(p)$ in Eq.\,\eqref{Spgen}: \emph{solid curves} -- continuum nonperturbative results \cite{Bhagwat:2003vw, Bhagwat:2006tu}; ``data'' -- numerical simulations of lQCD \protect\cite{Bowman:2005vx}.
The current-quark of perturbative QCD evolves into a constituent-quark as its momentum becomes smaller.  The constituent-quark mass arises from a cloud of low-momentum gluons attaching themselves to the current-quark.  This is DCSB, the essentially nonperturbative effect that generates a quark \emph{mass} \emph{from nothing}; namely, it occurs even in the absence of a Higgs mechanism.
The size of $M(0)$ is a measure of the magnitude of the QCD scale anomaly in $n=1$-point Schwinger functions \cite{Roberts:2016vyn}.
Moreover, experiments on $Q^2\in [0,12]\,$GeV$^2$ at the modern Thomas Jefferson National Accelerator Facility (JLab) will be sensitive to the momentum dependence of $M(p)$ within a domain that is here indicated approximately by the shaded region.
}
\end{figure}

\section{Dynamical Chiral Symmetry Breaking}
The emergence of a gluon mass-scale in the Standard Model drives an enormous array of phenomena.  Crucial amongst them is dynamical chiral symmetry breaking (DCSB), which is most readily apparent in the dressed-quark propagator:
\begin{equation}
\label{Spgen}
S(p) = 1/[i \gamma\cdot p A(p^2) + B(p^2)] = Z(p^2)/[i\gamma\cdot p + M(p^2)]\,.
\end{equation}
$S(p)$ can be predicted in QCD using nonperturbative continuum and lattice techniques.  The quantity $M(p^2)$ in Eq.\,\eqref{Spgen} is the dressed-quark mass-function, whose computed behaviour is depicted and explained in Fig.\,\ref{gluoncloud}.  Evidently, even in the absence of a Higgs mechanism, when current-masses vanish, quarks acquire a running mass whose value in the infrared is approximately $m_p/3$.  This is the scale required to support the constituent quark model and all its successes.  It follows that DCSB can be identified as the source for more than 98\% of the visible mass in the Universe, \emph{viz}.\ DCSB is fundamentally connected with the \emph{origin of mass from nothing}.

One must insist that chiral symmetry breaking in the absence of a Higgs mechanism is ``dynamical,'' as distinct from spontaneous, because nothing is added to QCD in order to effect this remarkable outcome and there is no change of variables in ${\mathpzc L}_{\rm QCD}$ that will make it apparent.  Instead, through the act of quantising the classical chromodynamics of massless gluons and quarks, a large mass-scale is generated in both the gauge and matter sectors.

DCSB is empirically revealed very clearly in properties of the pion.  In fact \cite{Roberts:2016vyn}, the key to understanding Eq.\,\eqref{EPTpion} is a set of Goldberger-Treiman-like (GT) relations \cite{Qin:2014vya, Binosi:2016rxz}, the best known of which states:
\begin{equation}
m \simeq 0 \; \left| \; f_\pi E_\pi(k;0) = B(k^2)
\right. ,
\end{equation}
where $E_\pi$ is the leading piece of the pion's Bethe-Salpeter amplitude (wave function), and $B$ is the scalar piece of the dressed-quark self-energy, Eq.\,\eqref{Spgen}. This equation is exact in chiral QCD and expresses the fact that the Nambu-Goldstone theorem is fundamentally an expression of equivalence between the quark one-body problem and the two-body bound-state problem in QCD's color-singlet flavor-nonsinglet pseudoscalar channel. Consequently and enigmatically:\\[0.5ex]
\hspace*{1em}\parbox[t]{0.44\textwidth}{\emph{The properties of the nearly-massless pion are the cleanest expression of the mechanism that is responsible for (almost) all the visible mass in the Universe}.}

\smallskip

\hspace*{-\parindent}The manner by which this can be exploited at modern and planned facilities, in order to chart, \emph{inter alia}, the origin and distribution of mass, is canvassed elsewhere \cite{Horn:2016rip, Denisov:2018unj, Aguilar:2019teb}.

With the GT relations in hand, one can construct an algebraic proof \cite{Qin:2014vya, Binosi:2016rxz}, that at any and each order in a symmetry-preserving truncation of those equations in quantum field theory necessary to describe a pseudoscalar bound state, there is a precise cancellation between the mass-generating effect of dressing the valence-quark and -antiquark which constitute the system and the attraction generated by the interactions between them, \emph{i.e}.
\begin{align}
M^{\rm dressed}_{\rm quark} + M^{\rm dressed}_{\rm antiquark}+ U^{\rm dressed}_{\rm quark-antiquark\;interaction} \stackrel{\rm chiral\;limit}{\equiv} 0\,.
\label{EasyOne}
\end{align}
This guarantees the ``disappearance'' of the scale anomaly in the chiral-limit pion.
An analogy with quantum mechanics thus arises: the mass of a QCD bound-state is the sum of the mass-scales characteristic of the constituents plus a (negative and sometimes large) binding energy.

Since QCD's interactions are universal, similar cancellations must take place within the proton. However, in the proton channel there is no symmetry that requires the cancellations to be complete. Hence, the proton's mass has a value that is typical of the magnitude of scale breaking in QCD's one body sectors, \emph{viz}.\ the dressed-gluon and -quark mass-scales.

The perspective just described may be called the ``DCSB paradigm''.  It provides a basis for understanding why the mass-scale for strong interactions is vastly different to that of electromagnetism, why the proton mass expresses that scale, and why the pion is nevertheless unnaturally light.  In this picture, no significant mass-scale is possible in QCD unless one of commensurate size is expressed in the dressed-propagators of gluons and quarks.

\section{Faddeev Equation for Baryons}
\label{FEBaryons}
The Faddeev equation was introduced almost sixty years ago \cite{Faddeev:1960su}.  It treats the quantum mechanical problem of three-bodies interacting via pairwise potentials by reducing it to a sum of three terms, each of which describes a solvable scattering problem in distinct two-body subsystems.  An analogous approach to the three-valence-quark (baryon) bound-state problem in quantum chromodynamics (QCD) is
explained in Refs.\,\cite{Cahill:1988dx, Reinhardt:1989rw, Efimov:1990uz}.  In this case, owing to DCSB  and the importance of symmetries \cite{Binosi:2016rxz}, a Poincar\'e-covariant quantum field theory generalisation of the Faddeev equation is required.  Like the Bethe-Salpeter equation for mesons, it is natural to consider such a Faddeev equation as one of the tower of QCD's Dyson-Schwinger equations (DSEs), which are being used to develop a systematic continuum approach to the strong-interaction bound-state problem \cite{Roberts:2015lja, Eichmann:2016yit}.

\begin{figure}[!t]
\begin{center}
\includegraphics[clip,width=0.95\linewidth]{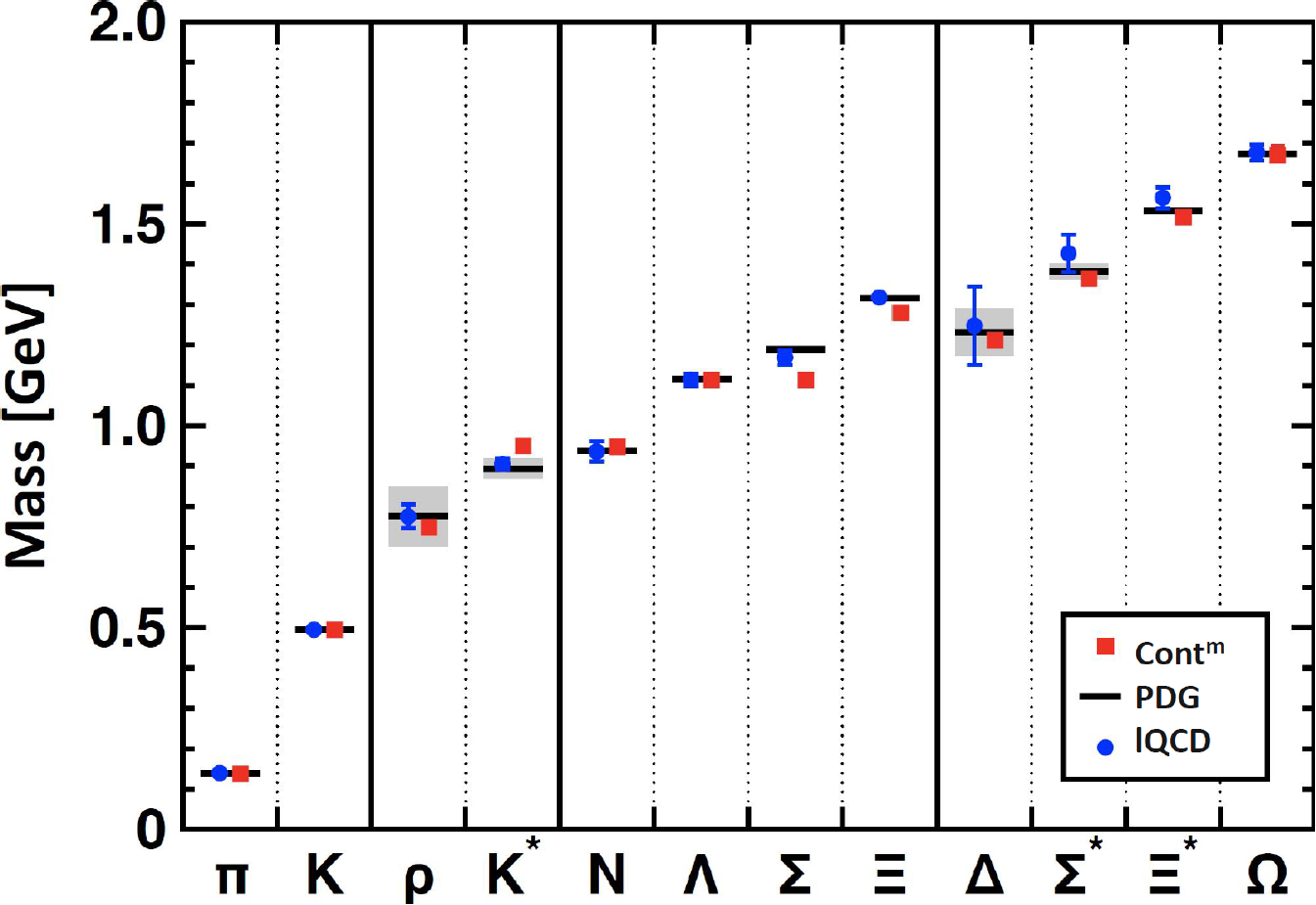}
\end{center}
\caption{\label{810MassComparison}
Masses of pseudoscalar and vector mesons, and ground-state positive-parity octet and decuplet baryons calculated using continuum (Cont$^{\rm m}$ -- squares, red) \cite{Qin:2019hgk} and lattice \cite{Durr:2008zz} methods in QCD compared with experiment (PDG -- black bars, with decay-widths of unstable states shaded in grey) \cite{Tanabashi:2018oca}.  The continuum study did not include isospin symmetry breaking effects, which are evidently small, as highlighted by the empirically determined $\Sigma$-$\Lambda$ mass difference.
}
\end{figure}

The first direct treatment of the Poincar\'e-covariant Faddeev equation for the nucleon is described in Ref.\,\cite{Eichmann:2009qa}.  Following that approach, Refs.\,\cite{Qin:2018dqp, Qin:2019hgk} calculated the spectrum of ground-state $J=1/2^+$, $3/2^+$ $(qq^\prime q^{\prime\prime})$-baryons, where $q, q^\prime, q^{\prime\prime} \in \{u,d,s,c,b\}$, their first positive-parity excitations and parity partners.  Introducing two parameters, to compensate for deficiencies of the leading-order truncation when used for excited hadrons, a description of the known spectrum of 39 such states was obtained, with a mean-absolute-relative-difference between calculation and experiment of $3.6\,(2.7)$\%.  This is exemplified in Fig.\,\ref{810MassComparison}.  The framework was subsequently used to predict the masses of 90 states not yet seen empirically.

Since solving the Faddeev equation yields Poincar\'e-covariant wave functions for all these systems, information about their internal structure is also revealed.  For instance, considering $J=3/2^+$ states, each has a complicated rest-frame angular-momentum profile.  The ground-states are primarily $S$-wave in character, but each possesses $P$-, $D$- and $F$-wave components: the $P$-wave is large in $\{u,d,s\}$-quark baryons. Turning to the first positive-parity excitations, each has a large $D$-wave component, which grows with increasing current-quark mass, but they otherwise exhibit features consistent with radial excitations.

Whilst a complete treatment of the Poincar\'e-covariant Faddeev equation is now possible using modern hardware, it remains a complex task.  Hence, for the development of insights in a wide array of baryon problems, it remains common to treat the equation in a quark-diquark approximation, where the diquark correlations are nonpointlike and dynamical \cite{Segovia:2015ufa}.  This amounts to a simplified treatment of the scattering problem in the two-body subchannels (as explained, \emph{e.g}.\ in Ref.\,\cite[Sec.\,II.A.2]{Hecht:2002ej}), which is founded on an observation that the same interaction which describes colour-singlet mesons also generates diquark correlations in the colour-antitriplet $(\bar 3)$ channel \cite{Cahill:1987qr, Maris:2002yu, Bi:2015ifa}.  Whilst the diquarks do not survive as asymptotic states, \emph{viz}.\ they are absent from the strong interaction spectrum \cite{Bender:1996bb, Bhagwat:2004hn}, the attraction between the quarks in the $\bar 3$ channel sustains a system in which two quarks are always correlated as a colour-$\bar 3$ pseudoparticle, and binding within the baryon is effected by the iterated exchange of roles between the bystander and diquark-participant quarks.  Contemporary studies indicate that diquark correlations are an important component of all baryons; and owing to the dynamical character of the diquarks, it is typically the lightest allowed diquark correlation which defines the most important component of a baryon's Faddeev amplitude \cite{Yin:2019bxe}.

\section{$\bf \Delta$ Baryons}
The quark+diquark picture of baryons has yielded a unified picture of the four lightest $(I,J^P) = (1/2,1/2^+)$ baryon isospin doublets \cite{Chen:2017pse}.  Turning to $(3/2,3/2^+)$ states with no net strangeness, the $\Delta^{+,0}$ baryons are conventionally viewed as the lightest isospin- and spin-flip excitations of the proton and neutron (nucleons, $N$), respectively.  Hence, since nucleons are the basic elements of all nuclei, developing a detailed understanding of the $\Delta$-baryons is of fundamental importance.  Without this, hadron physics remains at a level akin to atomic physics based only on knowledge of the hydrogen atom's ground state.

In this connection, given the simplicity of electromagnetic probes, it is worth using virtual photons to chart $\Delta$-resonance structure; and using intense, energetic electron-beams at the JLab, $\gamma^\ast p \to \Delta^+$ data are now available for $0 \leq Q^2 \lesssim 8\,$GeV$^2$ \cite{Aznauryan:2011ub, Aznauryan:2011qj, Aznauryan:2012ba}.
These data have stimulated much theoretical analysis and speculation about, \emph{inter alia}:
the role that resonance electroproduction experiments can play in exposing nonperturbative aspects of QCD, such as the nature of confinement and DCSB;
the relevance of perturbative QCD to processes involving moderate momentum transfers;
and hadron shape deformation.

Deformation can be characterised by a given hadron's spectroscopic quadrupole moment:
\begin{equation}
\label{deform}
Q = \frac{3 J_z^2 - J(J+1)}{(J+1)(2 J+3)} Q_0 ,
\end{equation}
where $Q_0$, the state's intrinsic deformation, is a measure of the $D$-wave component of its rest-frame-projected Faddeev wave function.  The difference between $Q$ and $Q_0$ expressed in Eq.\,\eqref{deform} represents the averaging of the nonspherical charge distribution due to its rotational motion as seen in the laboratory/rest-frame.  Plainly, $Q=0$ when $J=1/2$, \emph{i.e}.\ there is no measurable quadrupole moment for an isolated $J=1/2$ bound-state.

It is worth remarking here that just above the $\Delta$-baryon level lies the nucleon's first positive-parity excitation, \emph{viz}.\ the Roper resonance, $N(1440)\,1/2^+$.  Discovered in 1963 \cite{Roper:1964zza}, its characteristics were long the source of puzzlement because, \emph{e.g}.\ constituent-quark potential models typically (and erroneously) produce a spectrum in which this excitation lies above the first negative-parity state $N(1535)\,1/2^-$ \cite{Capstick:2000qj, Crede:2013sze, Giannini:2015zia}.  This has now changed following: acquisition and analysis of high-precision proton-target exclusive electroproduction data with single- and double-pion final states, on a large energy domain and with momentum-transfers out to $Q^2\approx 5\,$GeV$^2$; development of a dynamical reaction theory capable of simultaneously describing all partial waves extracted from available, reliable data; and formulation and application of a Poincar\'e covariant approach to the continuum bound-state problem in relativistic quantum field theory.  Today, it is widely accepted that the Roper is, at heart, the first radial excitation of the nucleon, consisting of a well-defined dressed-quark core that is augmented by a meson cloud, which both reduces the Roper's core mass by approximately 20\% and contributes materially to the electroproduction form factors at low-$Q^2$ \cite{Burkert:2017djo, Chen:2018nsg}.

Notably, a similar pattern of energy levels is found in the spectrum of $\Delta$-baryons, \emph{viz}.\ contradicting quark-model predictions,
the first positive-parity excitation, $\Delta(1600)\,3/2^+$, lies below the negative parity $\Delta(1700)\,3/2^-$, with the splitting being roughly the same as that in the nucleon sector.
This being so and given the Roper-resonance example, it is likely that elucidating the nature of the $\Delta(1600)$-baryon will require both (\emph{i}) data on its electroproduction form factors which extends well beyond the meson-cloud domain and (\emph{ii}) predictions for these form factors to compare with that data.
The data exist \cite{Trivedi:2018rgo, Burkert:2019opk}; and can be analysed with this aim understood.   This is especially important now that theoretical predictions are available \cite{Lu:2019bjs}.

The Poincar\'e-covariant wave functions for the $\Delta(1232)$ and $\Delta(1600)$ are depicted in Ref.\,\cite[Fig.\,2]{Lu:2019bjs}, from which it is plain that, in their rest-frames, both the ground-state and first positive-parity excitation are primarily $S$-wave in character.  Notably, the $\Delta(1232)$ mass is almost insensitive to non-$S$-wave components; and its quadrupole moment is large in magnitude and negative, indicating oblate intrinsic deformation.
Turning to the $\Delta(1600)$ and considering its mass, $P$-wave components generate noticeable repulsion, $D$-waves produce some attraction, and $F$-waves can be neglected.  Interestingly, in this quark+diquark approach, too, when comparing the ground state with the first positive-parity excitation, some of the $S$-wave strength is shifted into $P$- and $D$-wave contributions just as in the direct Faddeev equation solution described in Sec.\,\ref{FEBaryons}.  The intrinsic quadrupole moment of the $\Delta(1600)$ is just 45\% of that of the $\Delta(1232)$.  By this measure, the $\Delta(1600)$ is still oblate, but possesses less quadrupole deformation than the $\Delta(1232)$.

\section{$\mathbf{\gamma^\ast + p \to \Delta(1232)}$, $\mathbf{\Delta(1600)}$ }
Regarding the $\gamma^\ast p \to \Delta^+(1232)$ transition, on $Q^2 \gtrsim 0.5 m_p^2$, \emph{i.e}.\ outside the meson cloud domain for this process, the magnetic dipole and Coulomb quadrupole form factors reported in Ref.\,\cite{Lu:2019bjs} agree well with available data.  Consistent with the data, too, the electric quadrupole form factor is very small in magnitude; hence, it is particularly sensitive to the diquark content and quark-diquark angular-momentum structure of the baryons involved, and also to meson-baryon final-state-interactions (MB\,FSIs) on a larger domain than the other form factors.  These remarks are supported by the following observations: the role played by higher partial waves in the wave functions increases with momentum transfer (something also observed in meson form factors), here generating destructive interference; agreement with data on $G_M^\ast$ is impossible without the higher partial waves; and the effect of such components is very large in $G_E^\ast$, with the complete result for $G_E^\ast$ exhibiting a zero at $Q^2 \approx 4 m_p^2$, which is absent in the S-wave-only result(s).

\begin{figure}[t]
\centerline{%
\includegraphics[clip, width=0.42\textwidth]{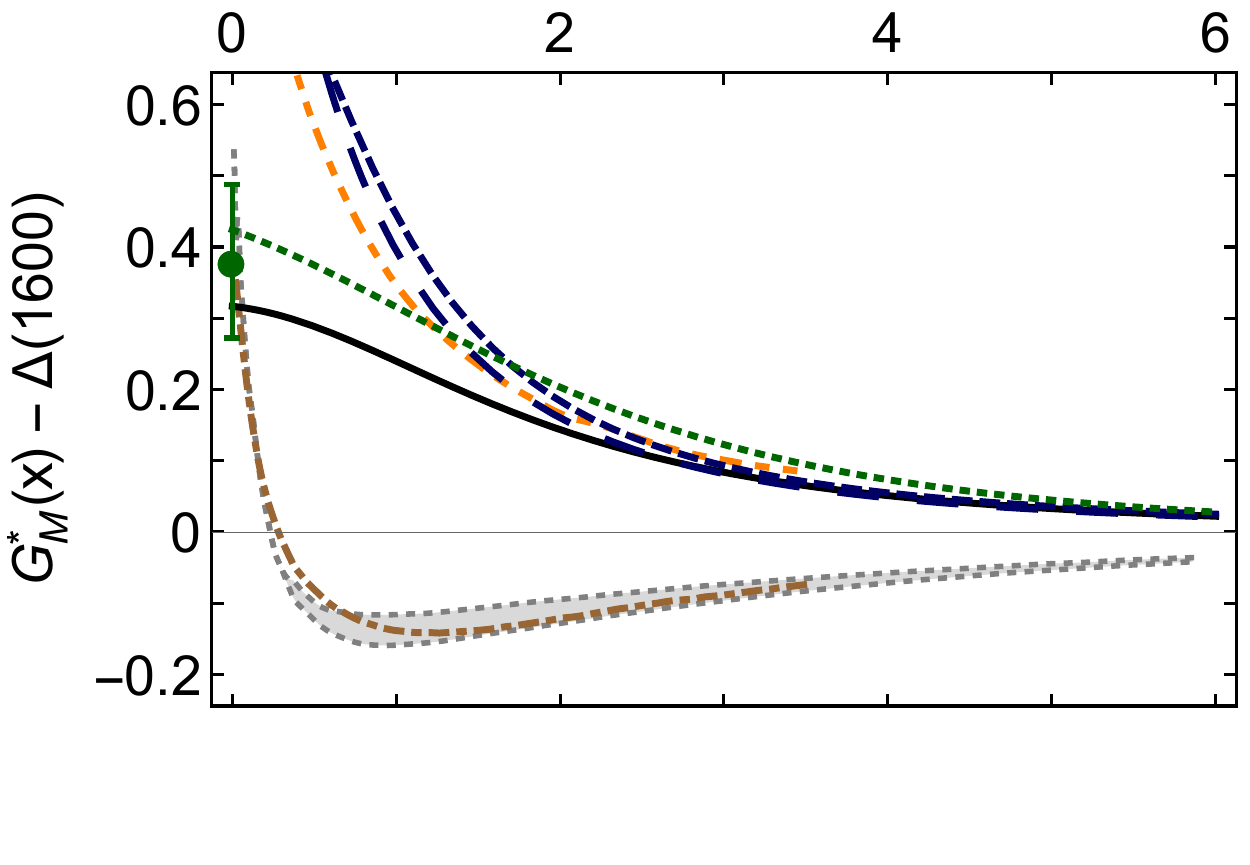}}
\vspace*{-8ex}

\centerline{%
\includegraphics[clip, width=0.42\textwidth]{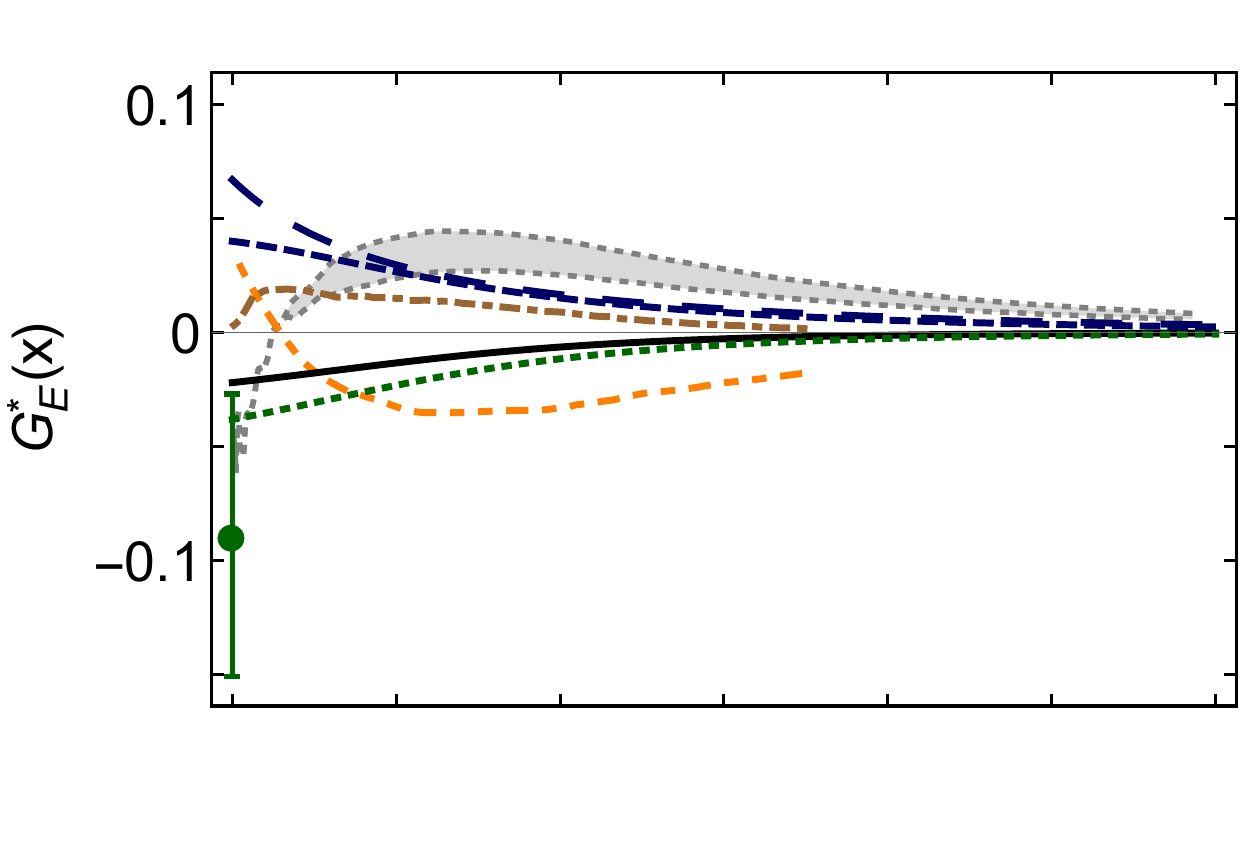}}
\vspace*{-8ex}

\centerline{%
\includegraphics[clip, width=0.42\textwidth]{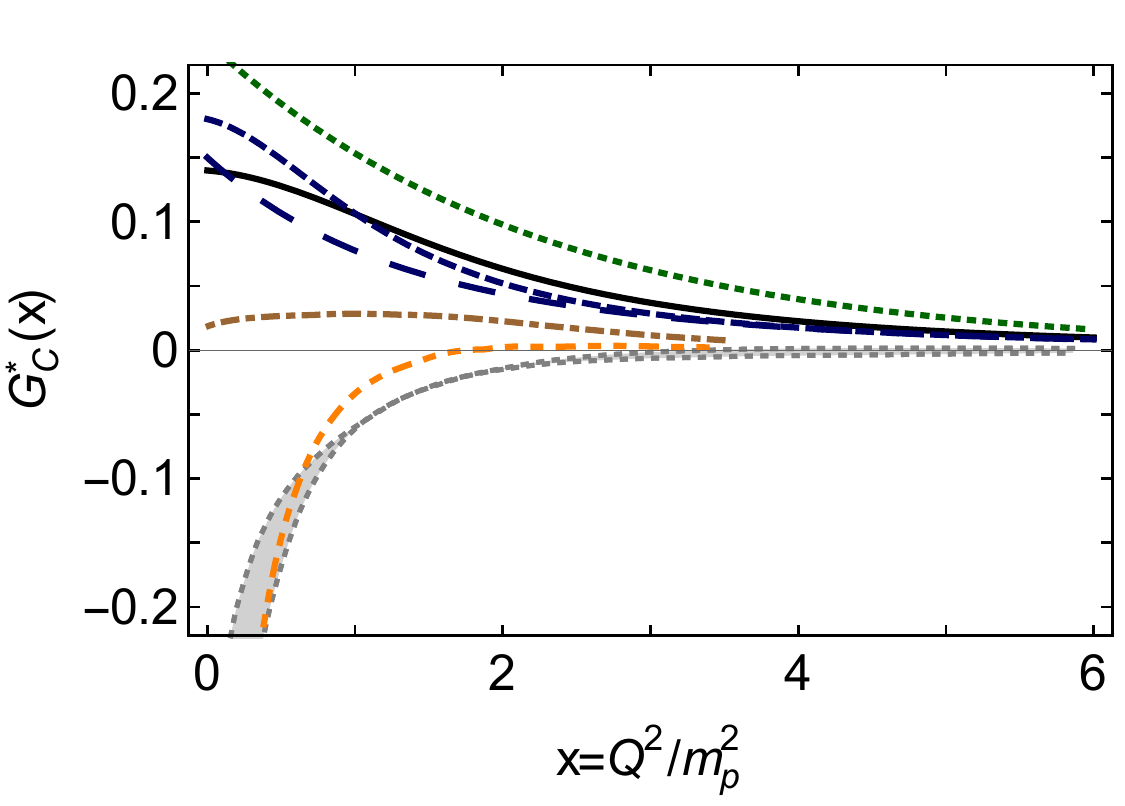}}
\caption{\label{D1600TFFs}
\emph{Top panel} -- Magnetic dipole $\gamma^\ast p\to \Delta^+(1600)$ transition form factor;
\emph{middle} -- electric quadrupole; and
\emph{bottom}:  Coulomb quadrupole.
Data from Ref.\,\cite{Tanabashi:2018oca}; and the conventions of Ref.\,\cite{Jones:1972ky} are employed.
All panels:
solid [black] curve, complete result;
long-dashed [blue] curve, result obtained when $\Delta(1600)$ is reduced to $S$-wave state;
dashed [blue] curve, both the proton and $\Delta(1600)$ are reduced to $S$-wave states;
dotted [green] curve, obtained by enhancing proton's axial-vector diquark content;
shaded [grey] band, light-front relativistic Hamiltonian dynamics (LFRHD) \cite{Capstick:1994ne};
dot-dashed [brown] curve, light-front relativistic quark model (LFRQM) with unmixed wave functions \cite{Aznauryan:2015zta};
and dot-dot-dashed [orange] curve, LFRQM with configuration mixing \cite{Aznauryan:2016wwmO}.
}
\end{figure}

Drawn from Ref.\,\cite{Lu:2019bjs}, predictions for the $\gamma^\ast p\to \Delta^+(1600)$ transition form factors are displayed in Fig.\,\ref{D1600TFFs}.  Empirical results are here only available at the real-photon point: $G_M^\ast(Q^2=0)$, $G_E^\ast(Q^2=0)$.  Evidently, the quark model results -- [shaded grey band] \cite{Capstick:1994ne}, dot-dashed (brown) curve \cite{Aznauryan:2015zta} and dot-dot-dashed [orange] curve \cite{Aznauryan:2016wwmO}) -- are very sensitive to the wave functions employed for the initial and final states.  Furthermore, inclusion of relativistic effects has a sizeable impact on transitions to positive-parity excited states \cite{Capstick:1994ne}.

The quark+diquark Faddeev equation prediction is the solid [black] curve in each panel of Fig.\,\ref{D1600TFFs}.  In this instance, every transition form factor is of unique sign on the domain displayed.  Notably, the mismatches with the empirical results for $G_M^\ast(0)$, $G_E^\ast(0)$ are commensurate in relative sizes with those in the $\Delta(1232)$ case, suggesting that MB\,FSIs are of similar importance in both channels.

Axial-vector diquark contributions interfere constructively with MB\,FSIs \cite{Hecht:2002ej}; hence, regarding form factors, one can mimic some meson-cloud effects by modifying the axial-vector diquark content of the participating hadrons.  Accordingly, to illustrate the potential impact of MB\,FSIs, the transition form factors were also computed using an enhanced axial-vector diquark content in the proton.  This was achieved by setting $m_{1^+} = m_{0^+} = 0.85\,$GeV, values with which the proton's mass is practically unchanged.  The procedure produced the dotted [green] curves in Fig.\,\ref{D1600TFFs}; better aligning the $x\simeq 0$ results with experiment and suggesting thereby that MB\,FSIs will improve the Faddeev equation predictions.

The short-dashed [blue] curve in Fig.\,\ref{D1600TFFs} is the result obtained when only rest-frame $S$-wave components are retained in the wave functions of the proton and $\Delta(1600)$; and the long-dashed [blue] curve is that computed with a complete proton wave function and a $S$-wave-projected $\Delta(1600)$.  Once again, the higher partial-waves have a visible impact on all form factors, with $G_E^\ast$ being most affected: the higher waves produce a change in sign.
This reemphasises one of the conclusions from the quark model studies, \emph{viz}.\ data on the $\gamma^\ast p\to \Delta^+(1600)$ transition form factors will be sensitive to the structure of the $\Delta^+(1600)$.

It is interesting to observe that whilst all $\gamma^\ast p \to \Delta^+(1232)$ transition form factors are larger in magnitude than those for $\gamma^\ast p \to \Delta^+(1600)$ in some neighbourhood of $Q^2=0$, this ordering is reversed on $Q^2 \gtrsim 2 m_p^2$.  One can thus argue that the $\gamma^\ast p \to \Delta^+(1600)$ transition is more localised in configuration space.

It is also notable that $R_{SM} \sim G_C^\ast/G_M^\ast$ is qualitatively similar for both $\gamma^\ast p \to \Delta^+$ transitions; but $R_{EM} \sim G_E^\ast/G_M^\ast $ is markedly different, being of opposite sign on $Q^2 \lesssim 4 m_p^2$ and uniformly larger in magnitude for the $\Delta(1600)$.  These observations again highlight the sensitivity of the electric quadrupole form factor to the degree of intrinsic deformation of the $\Delta$-baryons.

\section{Summary and Outlook}
\subsection{Particulars}
Poincar\'e-covariance entails that no realistic hadron wave function can be purely $S$-wave in character.  Hence, all $J\neq 0$ hadrons possess intrinsic deformation; and for baryons, such deformation can be exposed via measurements of nucleon-to-resonance transition form factors.  Of course, in relativistic quantum field theory, the magnitude and mixing between orbital angular momentum and spin is frame and renormalisation-scale dependent.  This is also true of the interpretations of these quantities because the natural degrees of freedom evolve with scale.  Consequently, even though Poincar\'e-invariant form factors are the same for all observers, their interpretation via subsystem degrees-of-freedom can change with frame and scale. All these remarks extend to deformation.  A close connection with quantum mechanics notions of probability distributions can be made via light-front projections of Poincar\'e-covariant wave functions; but unless the studied quantity is protected by symmetry, even conclusions drawn from such projections are scale dependent.

\subsection{Broader View}
$N^\ast$ physics has made great strides in this millennium.
For instance, continuum and lattice QCD approaches now appear to qualitatively confirm the constituent quark model spectrum; at least, they produce a spectrum of baryons that possesses a richness which cannot be explained by a two-body model; and the associated ``missing resonance'' problem is being addressed, with empirical gaps being filled.

Novel images of $N^\ast$ structure have been drawn through synergistic efforts between experiment and theory, with the availability of electroproduction data on a large $Q^2$ domain playing an especially important role.  A signature example here is the emerging understanding of the Roper resonance as the first radial excitation of the nucleon's dressed-quark core augmented by a meson cloud whose presence has a material impact on its low-$Q^2$ properties.   With this picture, the Roper becomes ``normal'', being much like the $\Delta$-baryon in many ways.

Such experiment-theory interplay is proving critical in validating modern QCD theory predictions for the momentum-dependence of the running-coupling, -masses, and other basic Schwinger functions that are computable using continuum and lattice methods.

The next decade will see completion of the analysis of existing electroproduction data available on $Q^2\in [0,5]\,$GeV$^2$ and the gathering of new results out to $Q^2 \approx 12\,$GeV$^2$.  The new large-$Q^2$ data will probe deeply into the quark-core domain of $N^\ast$ structure, enabling empirical flavour separation of transition form factor contributions and, thereby, stringent tests of theory predictions regarding the role of diquark correlations within baryons.  Furthermore, completed and new analyses of data on electroexcitation amplitudes of baryon parity partners will enable validation of predictions concerning the role of DCSB (emergent mass) in the spectrum and structure of nucleon resonances and help resolve questions relating to chiral symmetry restoration in highly excited systems.

Studies of these types, both ongoing and anticipated, will reveal facts about the running-coupling and -masses in QCD at length-scales associated with more-than 98\% of the observable mass of a typical hadron. They are therefore crucial in developing answers to some of the Standard Model's most fundamental questions; namely, what is mass, what is confinement, and how are they related?

\section*{Acknowledgements}
I am grateful to R.\,W.~Gothe for convening the session in which this perspective was presented.
The material contained herein is the product of collaborations with many people, to all of whom I am greatly indebted.
Related discussions can be found in contributions to this volume from, \emph{e.g}.\
V.~Burkert,
C.~Chen,
G.~Eichmann,
C.\,S.~Fischer,
R.\,W.~Gothe,
V.~Mokeev
and
J.~Rodr{\'{\i}}guez-Quintero.
Work partly supported by
Jiangsu Province \emph{Hundred Talents Plan for Professionals}.

%

\begin{thebibliography}{60}

\bibitem{Cornwall:1981zr}
J.M. Cornwall, Phys. Rev. D \textbf{26}, 1453 (1982)

\bibitem{Boucaud:2011ug}
P.~Boucaud, J.P. Leroy, A.~Le-Yaouanc, J.~Micheli, O.~Pene,
  J.~Rodr{\'i}guez-Quintero, Few Body Syst. \textbf{53}, 387 (2012)

\bibitem{Strauss:2012dg}
S.~Strauss, C.S. Fischer, C.~Kellermann, Phys. Rev. Lett. \textbf{109}, 252001
  (2012)

\bibitem{Binosi:2014aea}
D.~Binosi, L.~Chang, J.~Papavassiliou, C.D. Roberts, Phys. Lett. B
  \textbf{742}, 183 (2015)

\bibitem{Aguilar:2015bud}
A.C. Aguilar, D.~Binosi, J.~Papavassiliou, Front. Phys. China \textbf{11},
  111203 (2016)

\bibitem{Binosi:2016nme}
D.~Binosi, C.~Mezrag, J.~Papavassiliou, C.D. Roberts,
  J.~Rodr{\'i}guez-Quintero, Phys. Rev. D \textbf{96}, 054026 (2017)

\bibitem{Gao:2017uox}
F.~Gao, S.X. Qin, C.D. Roberts, J.~Rodr{\'{\i}}guez-Quintero, Phys. Rev. D
  \textbf{97}, 034010 (2018)

\bibitem{Cyrol:2017ewj}
A.K. Cyrol, M.~Mitter, J.M. Pawlowski, N.~Strodthoff, Phys. Rev. D \textbf{97},
  054006 (2018)

\bibitem{Rodriguez-Quintero:2018wma}
J.~Rodr{\'{\i}}guez-Quintero, D.~Binosi, C.~Mezrag, J.~Papavassiliou, C.D.
  Roberts, Few Body Syst. \textbf{59}, 121 (2018)

\bibitem{Binosi:2019ecz}
D.~Binosi, R.A. Tripolt (arXiv:1904.08172 [hep-ph]), {\emph{Spectral functions
  of confined particles}}

\bibitem{Accardi:2012qut}
A.~Accardi et~al., Eur. Phys. J. A \textbf{52}, 268 (2016)

\bibitem{NAP25171X}
\mbox{US National Academies}, \emph{An Assessment of U.S.-Based Electron-Ion
  Collider Science} (The National Academies Press, Washington, DC, 2018), ISBN
  978-0-309-47856-4

\bibitem{Brodsky:2015aia}
S.J. Brodsky, A.L. Deshpande, H.~Gao, R.D. McKeown, C.A. Meyer, Z.E. Meziani,
  R.G. Milner, J.W. Qiu, D.G. Richards, C.D. Roberts (aXiv:1502.05728
  [hep-ph]), {\emph{QCD and Hadron Physics}}

\bibitem{Deur:2008rf}
A.~Deur, V.~Burkert, J.P. Chen, W.~Korsch, Phys. Lett. B \textbf{665}, 349
  (2008)

\bibitem{Deur:2016tte}
A.~Deur, S.J. Brodsky, G.F. de~Teramond, Prog. Part. Nucl. Phys. \textbf{90}, 1
  (2016)

\bibitem{Bhagwat:2003vw}
M.S. Bhagwat, M.A. Pichowsky, C.D. Roberts, P.C. Tandy, Phys. Rev. C
  \textbf{68}, 015203 (2003)

\bibitem{Bhagwat:2006tu}
M.S. Bhagwat, P.C. Tandy, AIP Conf. Proc. \textbf{842}, 225 (2006)

\bibitem{Bowman:2005vx}
P.O. Bowman et~al., Phys. Rev. D \textbf{71}, 054507 (2005)

\bibitem{Roberts:2016vyn}
C.D. Roberts, Few Body Syst. \textbf{58}, 5 (2017)

\bibitem{Qin:2014vya}
S.X. Qin, C.D. Roberts, S.M. Schmidt, Phys. Lett. B \textbf{733}, 202 (2014)

\bibitem{Binosi:2016rxz}
D.~Binosi, L.~Chang, S.X. Qin, J.~Papavassiliou, C.D. Roberts, Phys. Rev. D
  \textbf{93}, 096010 (2016)

\bibitem{Horn:2016rip}
T.~Horn, C.D. Roberts, J. Phys. G. \textbf{43}, 073001 (2016)

\bibitem{Denisov:2018unj}
O.~Denisov et~al. (arXiv:1808.00848 [hep-ex]), {\emph{Letter of Intent (Draft
  2.0): A New QCD facility at the M2 beam line of the CERN SPS}}

\bibitem{Aguilar:2019teb}
A.C. Aguilar et~al., Eur. Phys. J. A \emph{in press}  (2019), {\emph{Pion and
  Kaon Structure at the Electron-Ion Collider} -- arXiv:1907.08218 [nucl-ex]}

\bibitem{Faddeev:1960su}
L.D. Faddeev, Sov. Phys. JETP \textbf{12}, 1014 (1961), [Zh. Eksp. Teor. Fiz.
  \textbf{39} (1960) 1459]

\bibitem{Cahill:1988dx}
R.T. Cahill, C.D. Roberts, J.~Praschifka, Austral. J. Phys. \textbf{42}, 129
  (1989)

\bibitem{Reinhardt:1989rw}
H.~Reinhardt, Phys. Lett. B \textbf{244}, 316 (1990)

\bibitem{Efimov:1990uz}
G.V. Efimov, M.A. Ivanov, V.E. Lyubovitskij, Z. Phys. C \textbf{47}, 583 (1990)

\bibitem{Roberts:2015lja}
C.D. Roberts, J. Phys. Conf. Ser. \textbf{706}, 022003 (2016)

\bibitem{Eichmann:2016yit}
G.~Eichmann, H.~Sanchis-Alepuz, R.~Williams, R.~Alkofer, C.S. Fischer, Prog.
  Part. Nucl. Phys. \textbf{91}, 1 (2016)

\bibitem{Eichmann:2009qa}
G.~Eichmann, R.~Alkofer, A.~Krassnigg, D.~Nicmorus, Phys. Rev. Lett.
  \textbf{104}, 201601 (2010)

\bibitem{Qin:2018dqp}
S.X. Qin, C.D. Roberts, S.M. Schmidt, Phys. Rev. D \textbf{97}, 114017 (2018)

\bibitem{Qin:2019hgk}
S.X. Qin, C.D. Roberts, S.M. Schmidt, Few Body Syst. \textbf{60}, 26 (2019)

\bibitem{Durr:2008zz}
S.~Durr et~al., Science \textbf{322}, 1224 (2008)

\bibitem{Tanabashi:2018oca}
M.~Tanabashi et~al., Phys. Rev. D \textbf{98}, 030001 (2018), {(\emph{Particle
  Data Group})}

\bibitem{Segovia:2015ufa}
J.~Segovia, C.D. Roberts, S.M. Schmidt, Phys. Lett. B \textbf{750}, 100 (2015)

\bibitem{Hecht:2002ej}
M.B. Hecht, C.D. Roberts, M.~Oettel, A.W. Thomas, S.M. Schmidt, P.C. Tandy,
  Phys. Rev. C \textbf{65}, 055204 (2002)

\bibitem{Cahill:1987qr}
R.T. Cahill, C.D. Roberts, J.~Praschifka, Phys. Rev. D \textbf{36}, 2804 (1987)

\bibitem{Maris:2002yu}
P.~Maris, Few Body Syst. \textbf{32}, 41 (2002)

\bibitem{Bi:2015ifa}
Y.~Bi, H.~Cai, Y.~Chen, M.~Gong, Z.~Liu, H.X. Qiao, Y.B. Yang, Chin. Phys. C
  \textbf{40}, 073106 (2016)

\bibitem{Bender:1996bb}
A.~Bender, C.D. Roberts, L.~von Smekal, Phys. Lett. B \textbf{380}, 7 (1996)

\bibitem{Bhagwat:2004hn}
M.S. Bhagwat, A.~H{\"o}ll, A.~Krassnigg, C.D. Roberts, P.C. Tandy, Phys. Rev. C
  \textbf{70}, 035205 (2004)

\bibitem{Yin:2019bxe}
P.L. Yin, C.~Chen, G.~Krein, C.D. Roberts, J.~Segovia, S.S. Xu, Phys. Rev. D
  \textbf{100}, 034008 (2019)

\bibitem{Chen:2017pse}
C.~Chen, B.~El-Bennich, C.D. Roberts, S.M. Schmidt, J.~Segovia, S.~Wan, Phys.
  Rev. D \textbf{97}, 034016 (2018)

\bibitem{Aznauryan:2011ub}
I.~Aznauryan, V.~Burkert, T.S. Lee, V.~Mokeev, J. Phys. Conf. Ser.
  \textbf{299}, 012008 (2011)

\bibitem{Aznauryan:2011qj}
I.~Aznauryan, V.~Burkert, Prog. Part. Nucl. Phys. \textbf{67}, 1 (2012)

\bibitem{Aznauryan:2012ba}
I.~Aznauryan, A.~Bashir, V.~Braun, S.~Brodsky, V.~Burkert et~al., Int. J. Mod.
  Phys. E \textbf{22}, 1330015 (2013)

\bibitem{Roper:1964zza}
L.D. Roper, Phys. Rev. Lett. \textbf{12}, 340 (1964)

\bibitem{Capstick:2000qj}
S.~Capstick, W.~Roberts, Prog. Part. Nucl. Phys. \textbf{45}, S241 (2000)

\bibitem{Crede:2013sze}
V.~Crede, W.~Roberts, Rept. Prog. Phys. \textbf{76}, 076301 (2013)

\bibitem{Giannini:2015zia}
M.M. Giannini, E.~Santopinto, Chin. J. Phys. \textbf{53}, 020301 (2015),
  \texttt{1501.03722}

\bibitem{Burkert:2017djo}
V.D. Burkert, C.D. Roberts, Rev. Mod. Phys. \textbf{91}, 011003 (2019)

\bibitem{Chen:2018nsg}
C.~Chen, Y.~Lu, D.~Binosi, C.D. Roberts, J.~Rodr{\'{\i}}guez-Quintero,
  J.~Segovia, Phys. Rev. D \textbf{99}, 034013 (2019)

\bibitem{Trivedi:2018rgo}
A.~Trivedi, Few Body Syst. \textbf{60}, 5 (2019)

\bibitem{Burkert:2019opk}
V.D. Burkert, V.I. Mokeev, B.S. Ishkhanov (arXiv:1901.09709 [nucl-ex]),
  {\emph{The nucleon resonance structure from exclusive $\pi^+\pi^-p$
  photo-/electroproduction off protons}}

\bibitem{Lu:2019bjs}
Y.~Lu, C.~Chen, Z.F. Cui, C.D. Roberts, S.M. Schmidt, J.~Segovia, H.S. Zong,
  Phys. Rev. D \textbf{100}, 034001 (2019)

\bibitem{Jones:1972ky}
H.F. Jones, M.D. Scadron, Annals Phys. \textbf{81}, 1 (1973)

\bibitem{Capstick:1994ne}
S.~Capstick, B.D. Keister, Phys. Rev. D \textbf{51}, 3598 (1995)

\bibitem{Aznauryan:2015zta}
I.G. Aznauryan, V.D. Burkert, Phys. Rev. C \textbf{92}, 035211 (2015)

\bibitem{Aznauryan:2016wwmO}
I.G. Aznauryan, V.D. Burkert (arXiv:1603.06692 [hep-ph]), {Configuration
  mixings and light-front relativistic quark model predictions for the
  electroexcitation of the \mbox{$\Delta(1232)3/2^+$}, \mbox{$N(1440)1/2^+$},
  and \mbox{$\Delta(1600)3/2^+$}}

\end{thebibliography}

\end{document}